\documentclass[pdftex]{article}

\usepackage[preprint,nonatbib]{neurips_2022}

\usepackage[utf8]{inputenc} % allow utf-8 input
\usepackage[T1]{fontenc}    % use 8-bit T1 fonts
\usepackage{hyperref}       % hyperlinks
\usepackage{tablefootnote}
\usepackage{siunitx}

\usepackage{url}            % simple URL typesetting
\usepackage{booktabs}       % professional-quality tables
\usepackage{amsfonts}       % blackboard math symbols
\usepackage{amsmath}       % blackboard math symbols
\usepackage{nicefrac}       % compact symbols for 1/2, etc.
\usepackage{microtype}      % microtypography
\usepackage{xcolor}         % colors
\usepackage[export]{adjustbox}
\usepackage[ampersand]{easylist}         % colors
\usepackage{lipsum}  
\usepackage{graphicx}
\usepackage{algorithm,algorithmicx,algpseudocode}
\usepackage{tabularx}
\usepackage[compact]{titlesec}
\titlespacing{\section}{0pt}{2ex}{1ex}
\titlespacing{\subsection}{0pt}{1ex}{0ex}
\titlespacing{\subsubsection}{0pt}{0.5ex}{0ex}

\title{Protein Structure and Sequence Generation with Equivariant Denoising Diffusion Probabilistic Models}

\author{%
  Namrata Anand  \\
  \texttt{namrata.anand2@gmail.com} \\
  \And 
  Tudor Achim  \\
  \texttt{tachim@cs.stanford.edu} \\
}

\begin{document}
\setlength{\abovedisplayskip}{0pt}
\setlength{\belowdisplayskip}{0pt}
\setlength{\abovedisplayshortskip}{0pt}
\setlength{\belowdisplayshortskip}{0pt}

\maketitle

\begin{abstract}
  Proteins are macromolecules that mediate a significant fraction of the cellular processes that underlie life.
  An important task in bioengineering is designing proteins with specific 3D structures and chemical properties which enable targeted functions.
  To this end, we introduce a generative model of both protein structure and sequence that can operate at significantly larger scales than previous molecular generative modeling approaches.
  The model is learned entirely from experimental data and conditions its generation on a compact specification of protein topology to produce a full-atom backbone configuration as well as sequence and side-chain predictions. 
  We demonstrate the quality of the model via qualitative and quantitative analysis of its samples. Videos of sampling trajectories are available at \url{https://nanand2.github.io/proteins}.
\end{abstract}

\section{Introduction}

Proteins are large macromolecules that play fundamental roles in nearly all cellular processes.
Two key scientific challenges related to these molecules are characterizing the set of all naturally-occurring proteins based on sequences collected at scale and designing new proteins whose structure and sequence achieve functional goals specified by the researcher.
Recently, AlphaFold2, a purely data-driven machine learning approach, has shown great progress in the forward problem of structure prediction \cite{jumper_highly_2021}.
Similarly, machine learning approaches have come to perform well for the sequence generation inverse problem \cite{anand_protein_2022,ingraham_generative_sequence_2019,fair-inversefolding-2022-sequences}.
However, for the task of structure generation, stochastic search algorithms based on handcrafted energy functions and heuristic sampling approaches are still in wide use \cite{rosetta_design_1,rosetta_design_2,rosetta_energy,rosetta-leaver-fay_scientific_2013}.

Data-driven generative modeling approaches have not yet had the same impact in the protein modeling setting as they have in the image generation setting because of several key differences. 
First, unlike images, proteins do not have a natural representation on a discretized grid that is amenable to straightforward application of existing generative models.
Interpreting the pairwise distance matrix of a protein's atoms as an image to be modeled with existing models has seen limited success because inconsistencies in the predictions lead to nontrivial errors when optimization routines are used to recover the final 3D structures \cite{anand_generative_2018}.
Second, unlike images, proteins have no natural canonical orientation. 
As a result, methods that are not rotationally invariant must account for this factor of variation directly in the model weights, which reduces the effective model capacity that can be dedicated to the structural variation of interest.
Finally, in protein generation, nontrivial errors in local \emph{or} global structure lead to implausible protein structures.

Previous work has made progress on different aspects of the problem.
Rotamer packing has benefited from machine learning approaches \cite{other_rotamer_paper,du2020energyrotamer,akpinaroglu2021improvedrotamer,mcpartlon2022attnpackerrotamer,anand_protein_2022}.
Machine learning has also made an impact on sequence design both in the case of conditioning on structural information \cite{anand_protein_2022,ingraham_generative_sequence_2019}, and without \cite{ferruz2022deepsequence,castro2022guidedsequence,strokach2020fastsequence,strokach2022deepsequence,ferruz2022towardssequence,gao2022alphadesignsequence,madani_progen_2020sequence,fair-inversefolding-2022-sequences}.
However, 3D molecular structure generation is a more challenging problem, and existing methods have either been limited to the generation of small molecules \cite{xu_geodiff_2021} or to large proteins in highly restricted settings with only one domain topology \cite{eguchi_ig-vae_nodate}.

\begin{figure}[h!]
  \centering
  \includegraphics[width=0.9\textwidth]{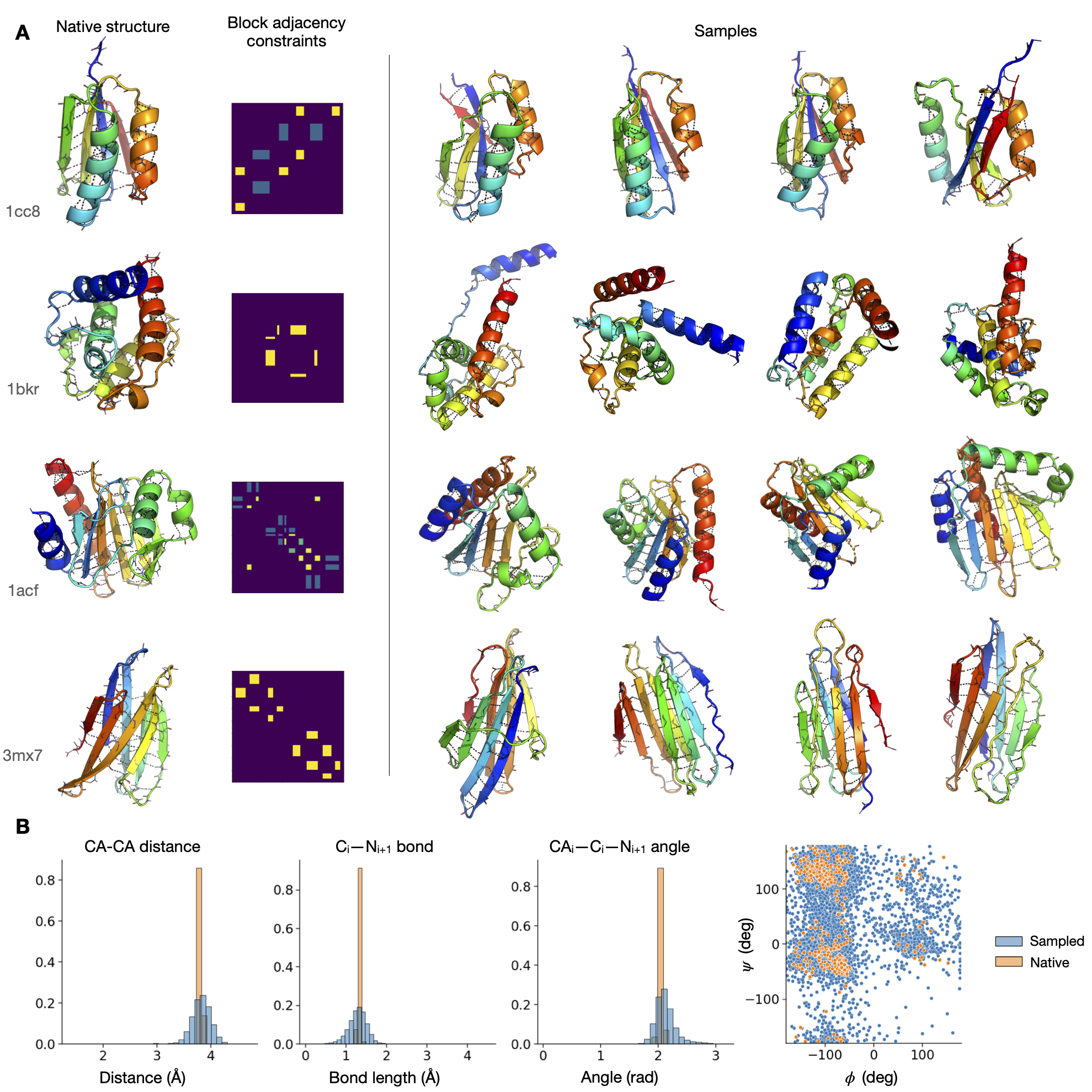}
  \caption{From-scratch protein generation. 
  Section A shows four different sampling scenarios.
  We show the block adjacencies in the middle and a test set structure matching the adjacencies on the left.
  On the right are four different samples from the model with no post-processing.
  The samples show a high degree of variability and excellent hydrogen bond patterns (the dashed lines) within helices and between beta sheets. Section B compares the distributions of bond lengths and angles and backbone torsions for the generated structures relative to native crystal structures. 
  }
    \label{fig:context-free}
\end{figure}

In this paper we present a new generative model that makes significant progress on closing this gap. 
We introduce a fully data-driven denoising diffusion probabilistic model (diffusion model) for protein structure, sequence, and rotamers that is able to generate highly realistic proteins across the full range of domains in the Protein DataBank (PDB) \cite{pdb}.
For comparison, protein macromolecules have approximately $100-1000\times$ the atom count of the small molecules addressed by previous molecular generative models, and the full set of domain types in the PDB numbers in the hundreds, in contrast to the single domain types addressed in previous work.
Our model is equivariant to rotations and translations using invariant point attention (IPA) modules introduced in \cite{jumper_highly_2021}.
To handle the diffusion of rotational frames of reference that are required for protein generation, we use a formulation that leverages an interpolation scheme well-suited to $SO(3)$.
For discrete sequence generation, we use an approach akin to masked language modeling that can be interpreted as diffusion in a discrete state space \cite{austin_structured_2021}.
Finally, to allow for interactive structure generation, we introduce a compact set of constraints that the model conditions on to generate proteins.
We show in Figure \ref{fig:context-free} and Section \ref{sec:experiments} that the model is able to generate high quality structures and sequences with nontrivial variety. 
To the best of our knowledge this is the first generative model that is capable of synthesizing physically plausible large protein structures and sequences across the full range of experimentally characterized protein domain topologies.

\section{Approach}

We first briefly describe the protein modeling problem and denoising diffusion probabilistic models.
Following that, we describe ways in which the diffusion training process is adapted to handling rotations as well as discrete (sequence) sub-problems.
Finally, we conclude the Approach section by introducing a compact encoding scheme for constraints that we use in conditional sampling of proteins, and summarize the training and sampling procedures.

\subsection{Preliminaries}

\paragraph{Proteins}

Proteins are macromolecules made up of chains of amino acids. The protein backbone consists of repeating atoms $N-C_\alpha-C-O$, with side-chains branching off the backbone from the $C_\alpha$ atom. Each group of backbone atoms with its associated side-chain is referred to as a residue. Interactions between the side-chains, the protein backbone, and
the environment give rise to local secondary structure elements – such as helices, strands, or loops – and to the ultimate tertiary structure of the protein. The 3D locations of the backbone and side-chain atoms fully describe the protein structure, and there are several priors that constrain the distribution of the atom locations.

First, the ordering of the backbone atoms from $N-$ to $C-$terminus is fixed, and bond lengths and angles between atoms on the backbone do not deviate much from average values. 
Second, the secondary structure of the protein is captured by the torsion angles $\phi, \psi$ of the backbone, which follow an established distribution known as the Ramachandran distribution. 
Alternatively, the $C_\alpha$ can be interpreted as forming a canonical orientation frame in relation to the $N$, $C$, and $O$ atoms, and the backbone atom positions as well as the torsion angles can be derived from these canonical frames. 
Third, atomic configurations of the 20 different amino acid side-chains can be described by some prefix of torsion angles $\chi_1,\chi_2,\chi_3,\chi_4$, which also follow amino-acid specific distributions. 
Although there are many experimentally- and theoretically-informed biophysical and statistical models for all of these quantities, we do not use them in training. 
In Section \ref{sec:experiments}, we measure how well our generative models recover these priors from the data.

Summarizing the above, assuming an $N$-residue protein, our goal is to learn a generative prior over the following variables:

\begin{easylist}[itemize]
& $x_{C_\alpha}^i \in \mathbb{R}^3$ for $i\in \{1, \ldots, N \}$, the 3D coordinates of the $C_\alpha$ backbone atoms. 
& $q^i\in SU(2)$, the unit quaternion defining the global rotation of the canonical frame centered at $x_{C_\alpha}^i$. Using $q^i$ and $x_{C_\alpha}^i$ we can recover the positions of the associated $N,C,O$ atoms in closed form and by extension the backbone torsion angles.
& $r^i\in \{1,\ldots,20\}$, the amino acid type of the $i^{th}$ residue.
& $\chi_1^i, \chi_2^i , \chi_3^i, \chi_4^i\in [-\pi, \pi)$, the four $\chi$ angles for the side-chain attached to the $i^{th}$ $C_\alpha$ atom. Note that some side-chains are made up of fewer atoms and thus have only a proper prefix of these angles.
\end{easylist}

\paragraph{Diffusion Models}

Diffusion models are a class of latent variable models that model the data generation process as iterative denoising of a random prior, with a specific parameterization of the approximate posterior distribution that can be interpreted as "diffusing" toward the fixed prior distribution \cite{sohl-dickstein_deep_2015}.
We briefly review the formulation below.
The data generation (reverse) process for a datapoint $x^0$ sampled from the data distribution $q(x^0)$ is defined recursively with a transition kernel $p_\theta$ and prior distribution $\pi$:

\begin{align}
    p_\theta\left(x^T\right) = \pi\left(x^T\right)
    \hspace{1.5cm}
    p_\theta\left(x^0\right) = \int_{x^{1:T}} p_\theta \left(x^T\right) \prod_{t=1}^T p_\theta \left(x^{t-1} | x^t\right)
\end{align}

The approximate posterior, referred to as the forward process, in the continuous case diffuses the datapoint $x^0$ toward the random prior:

\begin{align}
    q\left(x^{1:T}|x_0\right) = \prod_{t=1}^T \mathcal{N}\left(x^t; \sqrt{1-\beta_t}x^{t-1}, \beta_t I\right)
\end{align}

where the $\beta_t$ are chosen according to a fixed variance schedule. 
We use a neural network $\mu_\theta$ to parameterize the reverse transition kernel: $p_\theta(x^{t-1}|x^t)=\mathcal{N}(x^{t-1}; \mu_\theta(x^t, t), \sigma_t^2 I)$.
We obtain $\mu_\theta$ by minimizing the following variational bound during training, following \cite{ho_denoising_2020}:

\begin{align}
    L_{\text{simple}}(\theta) = \mathbb{E}_{t,x^0}\left[ \mathcal{L}_\text{FAPE}(x^0, \mu_\theta(x^t, t)) \right]
\end{align}

where $x^t$ is obtained by noising $x^0$ by $q$, and the rotationally invariant loss function $\mathcal{L}_\text{FAPE}$ is described in Section \ref{subsection:equivariant-training}. 
Finally, sampling relies on the learned $\mu_\theta$ to execute a reverse process with maps a sample from the prior distribution to a sample from the data distribution.

There are important differences between the image and protein generation settings which impact the architecture of $\mu_\theta$ as well as the training and sampling algorithms. 
The first we cover in the next section, and the second in Section \ref{subsec:putting-it-together} and Table \ref{table:diffusion-parameters}.

\subsection{Equivariant Diffusion Training}
\label{subsection:equivariant-training}

\paragraph{Diffusing Rotations}

Unlike coordinates, our rotation variables $q^i$ and $\chi_{1:4}^i$ do not live on Euclidean manifolds with flat geometry; therefore during training and sampling they cannot be diffused towards their prior distribution simply by randomly scaling and perturbing their encoding as is the case with coordinates.
To address such limitations, recent work has extended the diffusion framework to compact Riemannian manifolds \cite{de_bortoli_riemannian_2022}, which has been in turn adapted to modeling rotational diffusion as the repeated application of a heat kernel on a torus \cite{jing_torsional_2022}.
However, our experiments indicated that a simpler method suffices in practice.
For our prior distribution $\pi_q$ we sample uniform random rotations from $SU(2)$.
Next, instead of diffusing from $x^0$ towards $\pi_q$ with Brownian motion and thus modifying the reverse process to use an Euler-Maruyama sampler \cite{de_bortoli_riemannian_2022}, we \emph{interpolate} from $x^0$ to a sample $\epsilon\sim \pi_q$ based on the schedule of variances (see Table \ref{table:diffusion-parameters}).
We choose spherical linear interpolation ($\mathsf{SLERP}(x, y, \alpha)$, where we interpolate from $x$ to $y$ by a factor of $\alpha\in[0, 1]$) \cite{shoemake_animating_1985}.
For rotamer torsion angle diffusion (1D rotations) we use a uniform prior over $\mathcal{S}^1$ and interpolate on the unit circle for noising and sampling ("$\mathsf{Interp}")$.
These design choices have the desired effect of exposing the network to a similar distribution of random rotations both at training and test time, and our experiments demonstrate that they work well in practice.

\paragraph{Rotational Invariance}

As mentioned in the introduction, one key difference between images and proteins is that proteins have no canonical orientation.
As a result, we use an \emph{equivariant} transformer for our denoising model $\mu_\theta$.
The model takes as input an intermediate protein structure, $x^t$ and produces an estimate of the denoised ground truth structure $\hat{x}^0$.
We replace the standard attention mechanism in the transformer \cite{brown2020languagegpttransformer} with invariant point attention (IPA) as described in \cite{jumper_highly_2021}.
Put simply, IPA partitions node query and value features into 3-dimensional vectors and transforms them from the target node's reference frame into the a global reference frame before computing both attention weights and the output of the attention mechanism.
The output of the attention layer is invariant to the global orientation of the input protein, and thus the resulting corrections predicted by $\mu_\theta$ in the local coordinate frames of the $C_\alpha$s are equivariant. 

Training $\mu_\theta$ requires a loss function that can stably account for errors in all of the predictions of the generative model. Therefore we follow \cite{jumper_highly_2021} and use the frame-aligned point error (FAPE) loss. FAPE penalizes errors in rotation by aligning the predicted local transformation ($q^i, {x^i}_{C_\alpha}$) with the ground-truth local frame \emph{at each residue in turn} and computing the clamped squared distance between the ground-truth and predicted atoms. Because coordinate frames are aligned when computing loss, the training procedure is invariant to the orientation of protein structures in the dataset.

\subsection{Discrete Sequence Diffusion}
\label{subsection:discrete-diffusion}

We use an approach akin to a masked language model \cite{bert-transformer} to generate the sequences on top of the backbones, which can be interpreted as diffusion with an absorbing state \cite{austin_structured_2021}.
Concretely, we train the model by randomly masking a fraction of the residues, where the fraction is linearly interpolated in $[0, 1]$ during training as a function of $t$.
At test time, we run the reverse process by masking all residues at $t=T$, and iteratively sampling from a model whose input residues are masked independently with probability $t/T$, with $t$ stepping from $t=T$ to $t=0$ during sampling.

\subsection{Constraints}
\label{subsec:constraints}

Besides encoding a manifold on which relevant inverse problems can be solved, the main value of a generative model for protein structures and sequences is in allowing a researcher to specify simple, compact conditioning information encoding their desired structural properties, sample many valid protein configurations based on that, and iterate on the conditioning information until the desired results are obtained.
We introduce below one such constraint specification.

Given the secondary structure topology of a protein, the residues can be divided into contiguous adjacent blocks based on the secondary structure of the block (i.e. each block corresponds to either a helix, a beta sheet, or a loop of some length).
Furthermore, each pair of blocks can be considered to be adjacent or non-adjacent based on whether or not their closest atoms are within some distance threshold.
For paired beta sheets, besides adjacency we can specify whether the sheets are parallel or anti-parallel to each other.
Therefore, one way to compactly describe a protein is by specifying a number of residues $N$, then a tuple of numbers of length $B$ adding up to $N$ which indicate the block sizes, then a block secondary structure assignment $\{\text{helix}, \text{sheet}, \text{loop}\}^B$, and finally a symmetric block adjacency matrix in $\{0,1\}^{B\times B}$ together with a parallel/anti-parallel prior on each beta sheet pairing. In practice, we allow for the underspecification of the block adjacency matrix by dropping out block adjacencies during training and by not including any adjacency information for loop blocks by default.
This coarse encoding scheme for the topology prior does not overly constrain the model to produce just one structure; it allows for significant variation as seen in Section \ref{subsec:context-free}.

\subsection{Training and Sampling Summary}
\label{subsec:putting-it-together}

\begin{figure}[ht]
  \centering
    \includegraphics[width=0.8\textwidth]{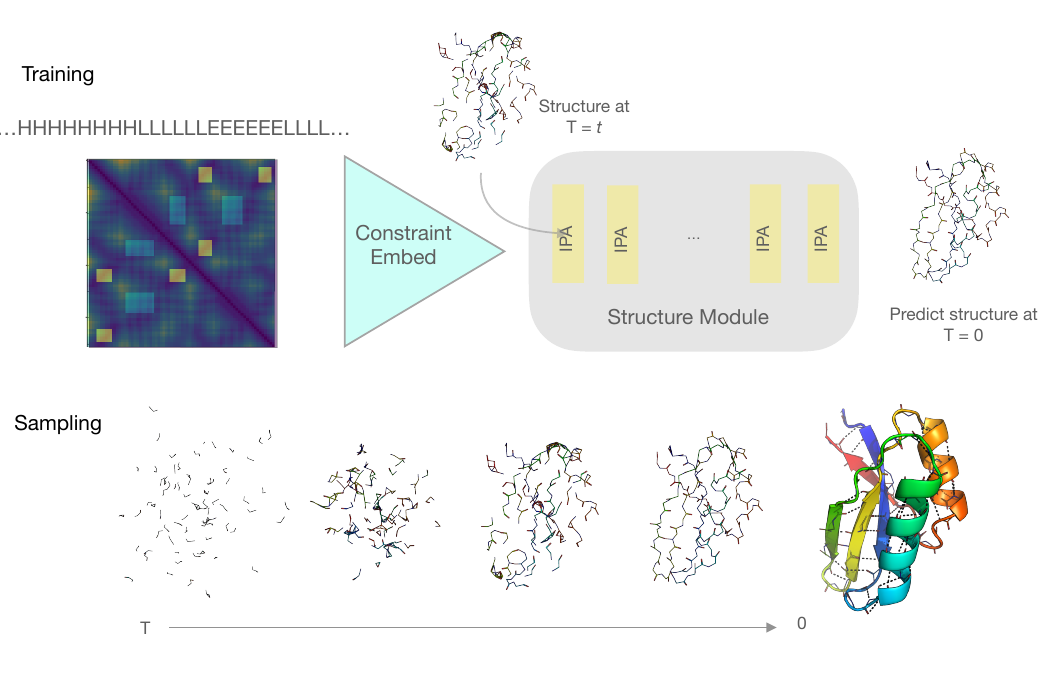}
  \caption{The model receives as input secondary structure and coarse constraints (shown here overlaid on the $C_\alpha$ distance matrix of the ground truth structure). 
  The constraints are used by the IPA modules during training to denoise the diffused input. 
  At test time, the model produces samples that adhere to the coarse constraints.
  }
  \label{fig:systemdiagram}
\end{figure}

We summarize the overall design choices here and in Figure \ref{fig:systemdiagram}.
The generative model conditions on a compact specification of constraints for a given protein, described in Section \ref{subsec:constraints}.
These constraints are embedded using a transformer with triangular self-attention to produce feature embeddings which are processed using Invariant Point Attention to produce updates to the translations, rotations, and residues in the local coordinate frames of the $C_\alpha$ atoms. 
These updates are used in training to compute rotationally-invariant losses, and in sampling to take steps toward the final structure.

In Table \ref{table:diffusion-parameters} we summarize for each variable the prior distribution, the approach used to interpolate between the data distribution and the noise distribution, and finally the method for taking a step at sampling time. 
To generate a structure, we sample a starting point from the prior distribution corresponding to $t=T$, and iteratively apply the update described in the "Sample Step" column in Table \ref{table:diffusion-parameters} for all variables for $t=T$ down to $t=1$.
The sample of the generative model is taken to be the value at $t=0$.

\begin{table}[ht]
  \caption{Diffusion Process Hyperparameters}
  \label{table:diffusion-parameters}
  \centering
  \begin{tabularx}{\textwidth}{llXX}
    Variable     & Prior Distribution ($\pi$)    & Training Noising (step $t$)& Sample Step (step $t$) \\
    \midrule
    $x_{C_\alpha}^i$ & $\mathcal{N}(0, I)$ &
    \multicolumn{2}{c}{Diffusion with $x_0$ prediction \cite{ho_denoising_2020} and cosine schedule \cite{dhariwal2021diffusion} }
    \\
    \hline
    $q^i$ & $\mathsf{Uniform}(SU(2))$ & $q^i_t = \mathsf{SLERP}(q^i_0, q^i_T, \alpha_t)$ & $q^i_{t-1}=\mathsf{SLERP}(\hat{q}^i_0, q^i_t, (1 -\alpha_t))$ \\
    \hline
    $\chi^i_{1:4}$ & $\mathsf{Uniform}(-\pi, \pi) $ & $\chi^i_t=\mathsf{Interp}(\chi_0, \chi_T, \gamma_t)$ & $\chi^i_{t-1}=\mathsf{Interp}(\chi^i_t, \hat{\chi}^i_0,  (1 -\gamma_t))$ \\
    \hline
    $r^i$ & Fully Masked & Mask each residue with probability $t/T$; predict and incur loss on masked. & Mask each residue with probability $t/T$; predict masked.  \\
    \bottomrule
  \end{tabularx}
\end{table}

\section{Experiments}
\label{sec:experiments}

We train our model on X-ray crystal structure data of CATH 4.2 S95 domains \cite{cath1,cath2} from the Protein Data Bank (PDB) \cite{pdb}.
We separate domains into train and test sets based on CATH topology classes, splitting classes into $\sim$ 95\% and 5\%, respectively (1374 and 78 classes, 53414 and 4372 domains each). 
This largely eliminates sequence and structural redundancy between the datasets, which enables evaluation of the approach's ability to generalize.

\paragraph{Constraint Embedding Model}
The secondary structure information is encoded via a 1D GPT-3-like architecture \cite{brown2020language}. Pairwise secondary structure embeddings and block adjacencies are then downsampled and passed through triangle multiplication and attention layers as in \cite{jumper_highly_2021}. The full details of the model architecture can be found in the Supplementary Material.

\paragraph{Diffusion Model $\mu_\theta$}

The diffusion model conditions on the output of the constraint network and the current structure and produces a guess for the final structure configuration. When predicting $\hat{x}_{C_\alpha}$ and $\hat{q}$, each IPA module produces an intermediate backbone update which we apply to the structure before computing the next round of IPA. The full details of the model architecture may be found in the Supplementary Material.

\paragraph{Training and Sampling}

During training we use the prior distributions and noising procedures described in Table \ref{table:diffusion-parameters}, sampling $t$ uniformly at random in $[1,T]$.
We use the AdamW optimizer and a cosine learning rate decay schedule \cite{adam,loshchilov2017decoupled,cosine-annealing}. 
The models are trained on single K80 and V100 GPUs on Google Cloud.

In the following three sections we use three separate models for structure ($x_{C_\alpha}$ and $q$), sequence ($r$), and rotamer ($\chi$) diffusion.
The structure model is trained using ground truth centered $x_{C_\alpha}$ coordinates that are scaled down $15\times$.
The sequence model is trained on ground truth structures, and the rotamer model is trained on ground truth structures and sequences.

We share results on context-free generation, protein completion, and sequence design and rotamer repacking. 
There is \emph{no post-processing} on the samples produced; all results are based on the raw output of the diffusion process at $T=0$.

\subsection{Context-free Generation}
\label{subsec:context-free}

We begin our analysis of our approach by assessing its performance on the task of synthesizing accurate 3D designs of proteins, relying just on the compact specification of the protein.
This task is difficult because the model must produce a physically plausible structure that also respects the coarse adjacency priors.
To assess the degree of generalization of the algorithm, we compare against native backbones from the test set which have CATH-defined topologies not seen by the model during training. 
We select four test case backbones that span the major CATH classes—all alpha, alpha–beta, and all-beta. 

We can see in Figure \ref{fig:context-free} that the model is able to produce structures that are highly variable and physically plausible.
In section A we show the test case native backbones and, for each one, we show the block adjacency and parallel/anti-parallel constraints as well as four high-fidelity samples from the model. 
The samples are of high quality, with intra-backbone hydrogen bonds forming within the helices as well as between the beta strands.
The beta sheets are especially challenging to synthesize because the local structure needs to be precisely correct for the bonds to form, which in turn imposes constraints on the global structure to support the positioning of the sheets. In Figure \ref{fig:random_backbones}, we show random samples for completeness. 

Quantitatively, the charts in section B indicate that the model has learned biophysical priors of proteins directly from the data distribution. 
The various bond lengths and angles show good histogram overlap between the native and sampled structures.
The distribution of generated backbone torsion angles are consistent with the Ramachandran distribution \cite{ramachandran_stereochemistry_1963}. 

\subsection{Inpainting and Controllable Generation} 

\begin{figure}[ht]
  \includegraphics[width=0.9\textwidth]{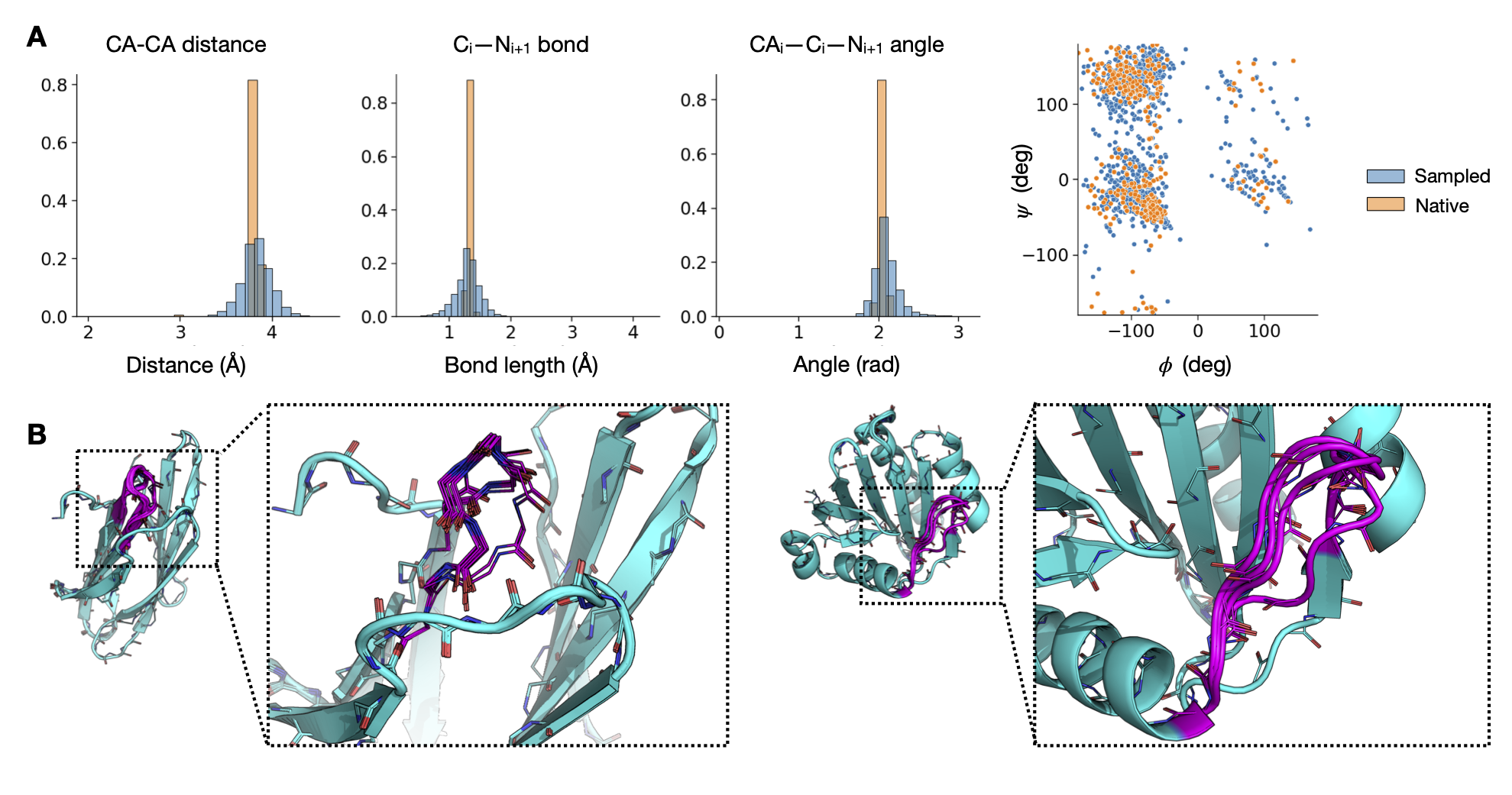}
  \centering
  \caption{Inpainting and loop design: Section A compares the distributions of bond lengths and angles and backbone torsions for the completed regions relative to the native structures. Section B shows examples of loop completions highlighted in purple. The image on the left highlights the model's ability to find discrete modes of the possible loop configurations.}
\label{fig:inpainting}
\end{figure}

We find that the model is also suitable for the task of completing existing proteins in novel ways.
For this task we train an additional model $\mu_\theta$ to condition on existing structures by holding parts of the structure fixed during training and executing the forward diffusion process on the complement of the fixed parts. For all residues that do not diffuse toward the prior, their position is held fixed at their ground truth positions during training and during sampling. Figure \ref{fig:inpainting} shows that the distribution of bond geometries for the inpainted regions is consistent with the corresponding distribution in the native structure.
We also see from the samples that the model can find discrete modes of the loop distribution at the atomic level. Random samples per test case are given in Figure \ref{fig:random_inpaint}.

\begin{figure}[ht]
  \includegraphics[width=0.9\textwidth]{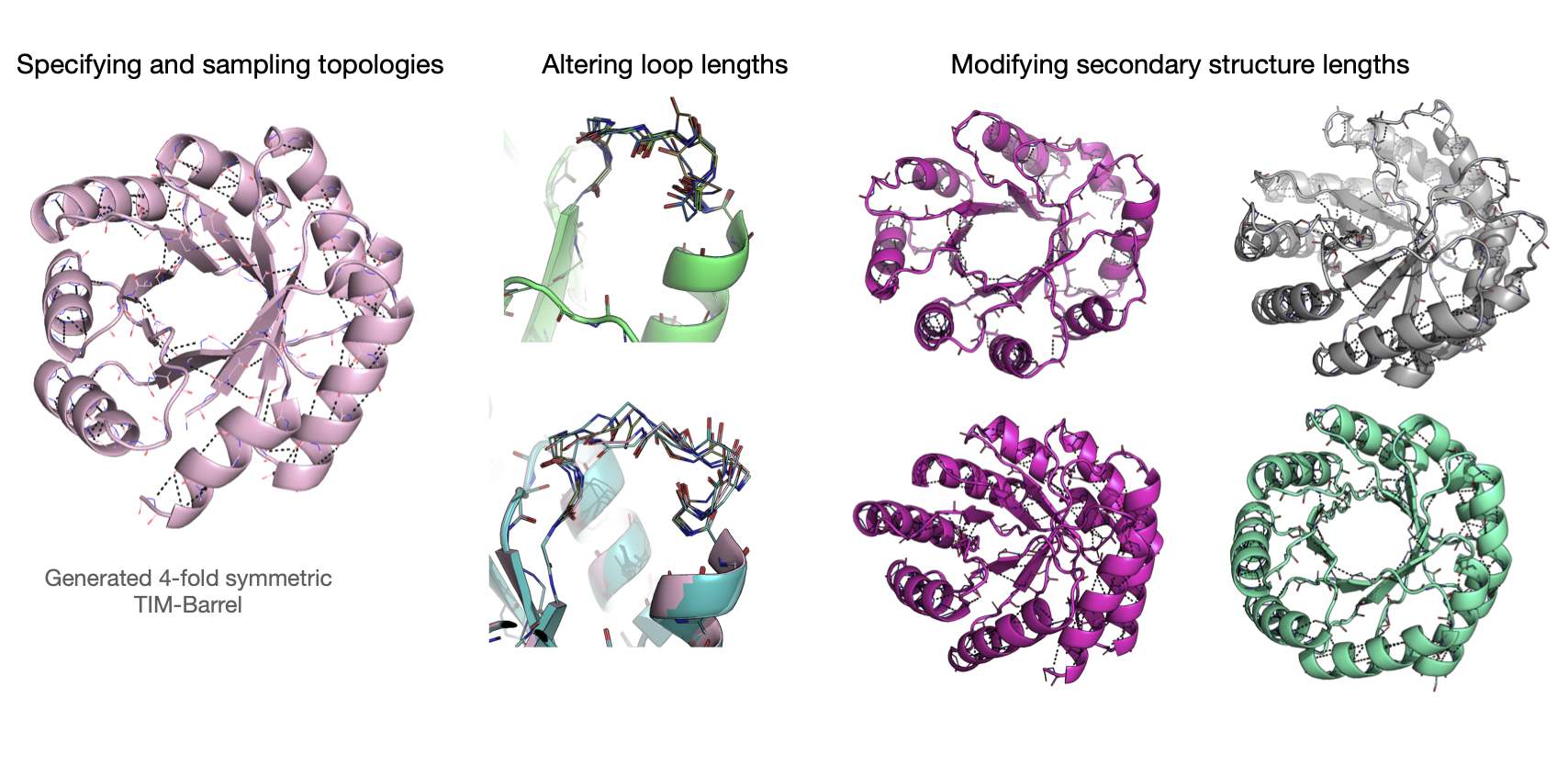}
  \centering
  \caption{Controllable generation: The model enables many modes of controllable generation of protein structure. Here, we highlight (A) new and/or idealized topology generation, (B) loop engineering, and (C) secondary structure modification. 
}
\label{fig:controllable-generation}
\end{figure}

We explore whether the model can go beyond sampling variants of topologies of existing proteins, to modifying the topologies themselves.
In this case, we use the same underlying $\mu_\theta$ model as for the inpainting case but at sampling time we modify the block adjacency conditioning information by simply modifying the lengths of underlying secondary structures for blocks. 
In Figure \ref{fig:controllable-generation}, we show how the model can be used to modify the structures in physically plausible ways -- generating idealized topologies, altering loop lengths, and modifying secondary structure lengths for a fixed topology.
We emphasize that these synthetic structures are distinct from the natural structures found in the PDB, which suggests that the model has encoded useful physical priors for use in sampling. In Figure \ref{fig:random_TIMs}, we show additional random TIM-barrel samples, and in Figure \ref{fig:modify_loop_len} we provide additional examples of model contextual design of variable-length loops.

\subsection{Sequence Design and Rotamer Packing} 

\begin{figure}[ht]
  \centering
    \includegraphics[width=0.8\textwidth]{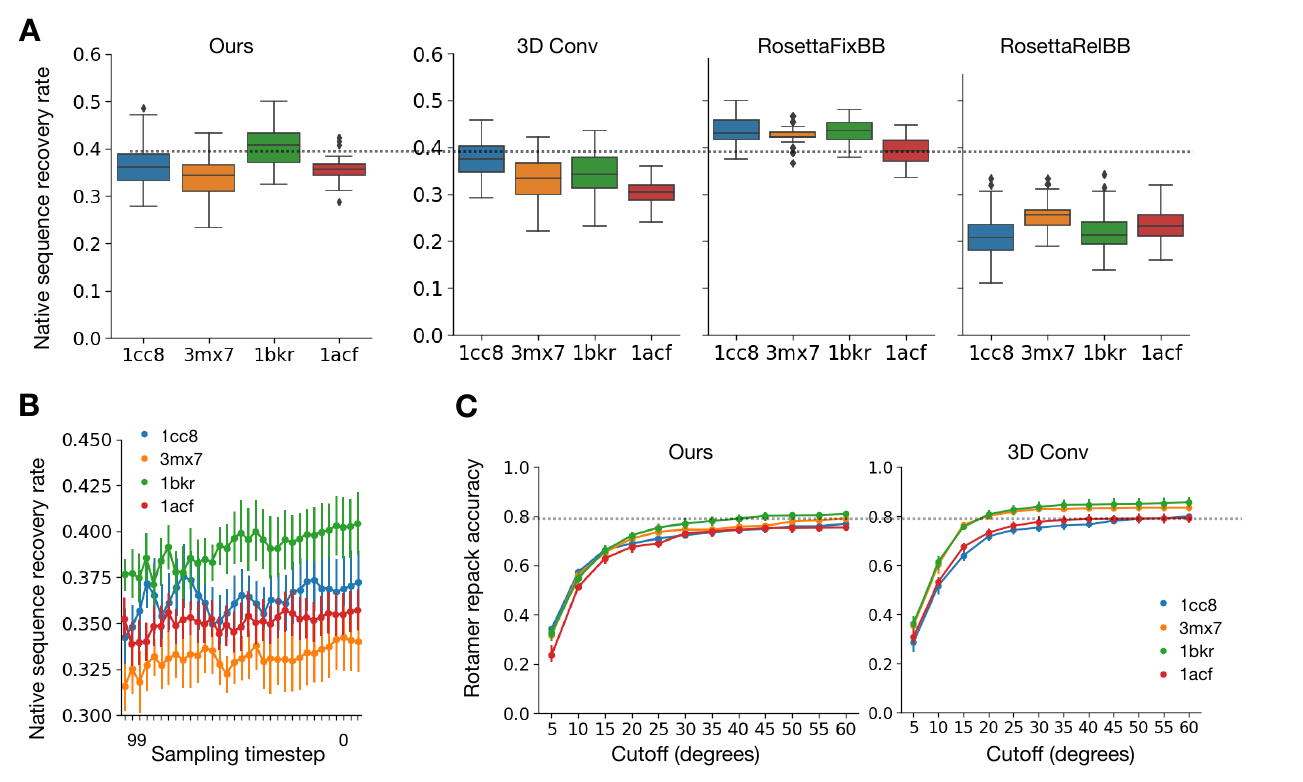}
  \caption{Sequence design and rotamer repacking. Section A reports native sequence recovery rates across 50 designs for test case structures. We reproduce design baselines reported in \cite{anand_protein_2022}. Section B shows the sequence recovery rate during the sampling trajectory, starting from predicting from all masked tokens. Section C shows the rotamer packing accuracy after $\chi$ diffusion as a function of degree cutoff, baselined against data reported in \cite{anand_protein_2022}. This approach to sequence design and rotamer packing is comparable to baselines and faster by an order of magnitude.}
\label{fig:sequence}
\end{figure}

We see similarly strong performance for sequence design and rotamer packing as for structure generation.
We measure the model's ability to recover ground truth sequences and rotamer configurations on native structures, because the physical variation in sampled structures implies a different set of optimal residues and rotamer configurations which cannot be compared to the ground truth directly.
The sequence recovery rates are compared across 50 sampled sequences, each starting from the native full-atom backbone with no side-chain information.
In Figure \ref{fig:sequence} we see that the model has comparable sequence recovery performance to baselines \cite{anand_protein_2022}.
The $\mathsf{3D Conv}$ baseline refers to a machine learning approach for sequence design and rotamer packing using 3D convolutions \cite{anand_protein_2022}. 
$\mathsf{RosettaFixBB}$ and $\mathsf{RosettaRelBB}$ are baselines using heuristic energy functions; $\mathsf{RosettaFixBB}$ holds the backbone fixed during sequence sampling, which is the same setting as our model, and $\mathsf{RosettaRelBB}$ allows it to vary slightly in a "relaxation" procedure \cite{rosetta-leaver-fay_scientific_2013}. 
The rotamer packing performance is comparable at the most stringent metric cutoffs (5 and 10 degrees).

\subsection{Joint Modeling}

In the previous sections we considered one model each for structure, sequence, and rotamer diffusion.
We now compare to a model trained to jointly diffuse structure and sequence concurrently.
Structure variables $x_{C_\alpha}$ and $q$ are diffused for the full $T_{structure}=1000$ steps with the diffusion training and sampling approaches described in Section \ref{subsec:putting-it-together}.
The sequence variables $r$ are diffused from $T_{sequence}=100$ to $T=0$, with an additional network that conditions on the output of the structure component of $\mu_\theta$ at each step.
That is, for a given $0\leq t\leq 100$, we perform masked prediction of the sequence using the schedule in Table \ref{table:diffusion-parameters}, conditioning on the prediction of $\hat{x}^0_{C_\alpha}$ and $\hat{q}^0$ from the structure network. 
Rotamer diffusion is then run on the sampled backbone and sequence.

\begin{figure}[ht]
  \centering
  \includegraphics[width=0.8\textwidth]{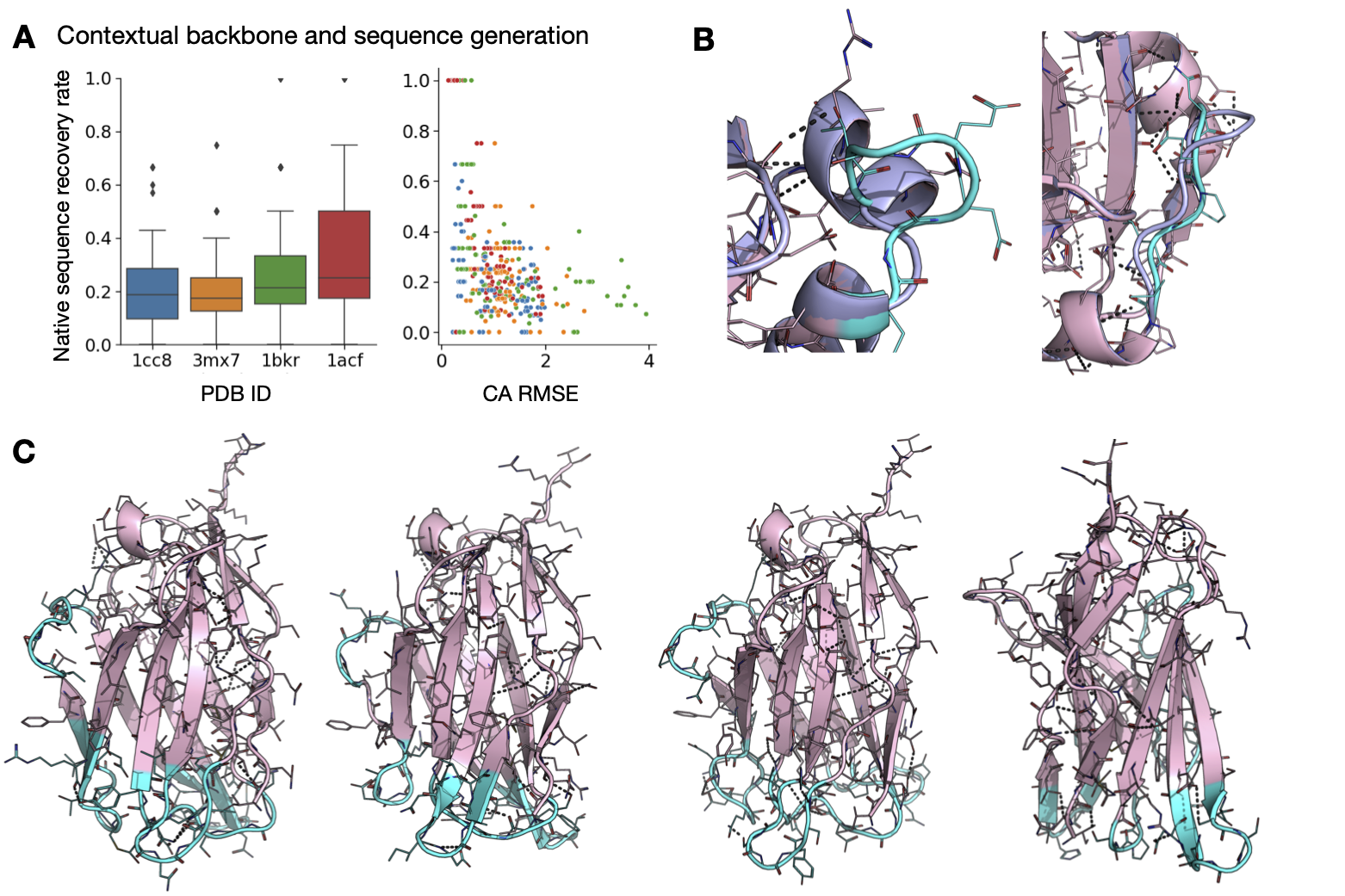}
  \caption{Contextual joint sampling of sequence and structure. For all cases shown, rotamers are packed with the rotamer diffusion model. Section A reports native sequence recovery rates and $C_\alpha$ RMSE after inpainting masked regions of test case proteins and sampling both backbone and sequence. While we do not expect perfect sequence recovery, we see that for some cases the model can nearly recover the native loop and sequence. Section B shows examples of model generated loops and sequences (cyan) with the native backbone (purple) for context. In Section C, given a fixed immunoglobulin backbone and sequence (pink), we sample variable-length loops and residues (cyan) jointly.}
  \label{fig:joint}
\end{figure}

In Figure \ref{fig:joint}a, we do contextual inpainting of both the backbone and sequence and find that the model is able to at times nearly recover the native solution both in terms of native sequence and backbone atom positions for inpainted regions. This type of a model enables us to do, for example, full-atom loop generation (Figure \ref{fig:joint}b) where we generate both the loop backbone and candidate sequence for the loop region jointly. This capability opens up the avenue to interesting engineering problems, such as immunoglobulin (Ig) loop design. Antibody variable Ig domains host highly variable CDR (complementarity-determining region) loops that allow them to selectively bind practically any target. In Figure \ref{fig:joint}c, we demonstrate how this type of generative model can be used to vary the CDR backbone loops and sequence jointly on a fixed Ig backbone.

Ultimately, we want to be able to sample jointly over backbone and sequence in a way that is self-consistent -- namely, that the generated sequence folds to the generated backbone structure. Although we can jointly sample structures and corresponding sequences from scratch with this approach, sample quality will likely improve once we allow the network to cross-condition on structure, sequence, and rotamers, rather than generate them sequentially, as in the current set-up. We leave exploration of models of this type to future work.

\section{Conclusions}

We have introduced a generative model of protein structures, sequences, and rotamers that can produce large protein domains that are both physically plausible and highly varied across domain types in the PDB.
To this end, we designed a compact constraint specification which our model conditions on to produce highly-varied proteins.
We demonstrated the model's performance both qualitatively and quantitatively using biophysical metrics and showed its potential for modifying existing proteins, from designing loops to varying the underlying topology. 
We concluded with an analysis of the model's ability to design sequences and pack rotamers, indicating its potential as a fully end-to-end tool for protein design.

There are many interesting areas for further exploration.
First, one may replace the supervised learning "recycling" procedure for predictions in AlphaFold2 with the diffusion formulation in this paper (or, equivalently, replace the Constraints conditioning information with the output of the Evoformer blocks from \cite{jumper_highly_2021}).
In predicting the structure of a protein there is often nontrivial aleatoric uncertainty, which arises from the fact that that there are often many conformations that the protein could adopt, of which we only observe one via crystallography.
Our model enables a simple way of quantifying uncertainty, via measurement of the spread of samples, which may be of interest to practitioners as an additional signal beyond the per-residue uncertainty quantification made available by AlphaFold2 \cite{jumper_highly_2021}.

Second, auxiliary energy functions could be used to interactively guide sampling for more fine-grained control over the sampling process.
The current constraint specification is intentionally compact to allow for easy specification as well as wide variance in the generated structures.
However, the gradients of simple energy functions (such as ones that penalize deviation from distance constraints) could guide more precise modifications during sampling.

Third, we anticipate natural applications of this type of model to problems in rational design and structure determination. 
Adaptations of these models could be effective for direct synthesis of proteins in protein-protein complexes. 
Moreover, we anticipate natural applications of this type of model in fitting proteins to Cryo-EM volumes. 
Current approaches typically use auto-regressive methods to iteratively fit structures and can face difficulties when the volumes are ambiguous and significant backtracking and search is necessary to correct for mistakes in the fitting process.
Our protein diffusion model, which forms the structure globally during sampling, may help mitigate these failure modes.

While it is highly beneficial for researchers to have access to better models that can inform the design of therapeutics, vaccines, and more, there are indeed risks associated with having powerful tools for protein design.
It may be the case that the accelerated rate of progress in computational methods brought on by the application of AI techniques makes it more difficult for the research to community to self-regulate in response to rapid changes in method capabilities.

\section*{Acknowledgements}

We thank Jonathan Ho for helpful discussion on diffusion models.

\clearpage
\bibliographystyle{plain}

\clearpage
\appendix

\section{Model details }

\subsection{Training}

Models are optimized with the AdamW optimizer with and a cosine learning rate decay schedule \cite{adam,loshchilov2017decoupled,cosine-annealing}. 
All models are trained on single K80 and V100 GPUs on Google Cloud. We use gradient accumulation to increase the effective batch sizes.

\subsection{Constraint embedding}

 Beta stands are considered adjacent if their minimum $C_\alpha$ pairwise distance is less than 5 $\si{\angstrom}$ during training. When considered adjacent, the orientation (parallel or anti-parallel) of beta-beta pairing is also given as an input. 
 The other pairs of secondary structure elements (beta-helix, helix-helix) are considered adjacent if their minimum $C_\alpha$ pairwise distance is less than 7 $\si{\angstrom}$. No adjacency information is encoded for loop blocks.

The constraint embedding network consists of a 5-layer GPT-3-like transformer followed by 25 layers of triangle attention blocks \cite{jumper_highly_2021}. 

\subsection{Diffusion decoders}

\paragraph{Structure diffusion}
The diffusion decoder model conditions on input constraints and intermediate noised protein structures and predict corrected protein structures. The model consists of layers of Invariant Point Attention (IPA) blocks \cite{jumper_highly_2021}, which encode the current structure and features, predict backbone update parameters, and then update the structure rotations and translations. 

 The structure diffusion network has 12 layers, 4 IPA heads, 4 query points per residue, 4 value points per residue, and is trained with subsampled backbones up to to 256 residues. The network is trained with $T=1000$, batch size of 160, and learning rate $10^{-3}$.

 \paragraph{Sequence diffusion}
 For sequence diffusion, no constraint embeddings are used, and no backbone updates are done in the forward pass. 
 
 The network has 15 layers, 4 IPA heads, 4 query points per residue, 4 value points per residue, and is trained with subsampled backbones up to to 128 residues. The network is trained with $T=100$, batch size 160, and learning rate $10^{-3}$.

  \paragraph{Rotamer diffusion}
 For rotamer diffusion, the model sees as an input the entire full-atom protein structure with diffused side-chains. No constraint embeddings are used, and no backbone updates are done in the forward pass. The network has 6 layers, 4 IPA heads, 4 query points per residue, 4 value points per residue, and is trained with subsampled backbones up to to 75 residues. The network is trained with $T =500$, batch size 100, and learning rate $5 \times 10^{-4}$.

 \paragraph{Structure inpainting}
 
 For structure inpainting, we keep the entire structure fixed except for a region that is masked. For each datapoint during training, we execute "block diffusion" with probability 0.6 and "contiguous diffusion" with probability 0.4. In block diffusion each loop is masked with probability 0.25, and the other blocks are masked with probability 0.05. In contiguous diffusion, we choose contiguous blocks at random to diffuse towards the prior with probability 0.03 for each starting residue and with length distributed uniformly between 1 and 15.

The structure inpainting model is fine-tuned from the structure diffusion model with $T=1000$, batch size 160, and learning rate $1 \times 10^{-3}$.

 \paragraph{Joint structure and sequence inpainting}
 
 For joint structure and sequence diffusion, the pretrained structure inpainting network is kept frozen, and the pretrained sequence diffusion network is finetuned on the predicted structures outputted by the structure diffusion network. The sequence model is trained with $T_{\text{sequence}}=100$, batch size 160, and learning rate $5 \times 10^{-4}$.

 \section{Experiment details}
 
 \subsection{TIM-barrel topology sampling}
 We specify TIM-barrel block adjacencies corresponding to a parallel beta-barrel surrounded by a ring of helices that are each adjacent to their immediate neighbors. We assume each quarter of the TIM-barrel to have the following secondary structure blocks in order: L-E-L-H-L-E-L-H-L (where L -- loop, E -- beta strand, H -- helix). We sample random lengths for these secondary structure blocks with reasonable selected lower and upper bounds, and repeat the sampled topologies 4$\times$ for the four-fold symmetric topology. We eliminate topologies longer than 256 residues. We then run the diffusion model on the sampled secondary structure string and block adjacencies to decode a range of TIM-barrel structures with varied underlying topologies.

  \subsection{Ig domain joint loop backbone and sequence sampling}
  
  We start with a random Ig domain from the train set with CATH ID 5tdoA01. We mask out the CDR loops, as well as an additional loop that might interact with the CDRs but is not natively hypervariable. We then sample random loop lengths for these regions. Note that the adjacency information does not change for the structure, as we do not encode any adjacency information for loop regions. Moreover, note that although CDR loops can have some structured regions, we ask the model to produce loops only. We run the joint structure and sequence model to generate new variable-length loop backbones as well as a sequence for the new loop regions. We then use the learned rotamer packer to pack the sequence of the entire structure, including the new loop side-chains.

\clearpage

\begin{figure}[ht]
  \centering
    \includegraphics[width=0.95\textwidth]{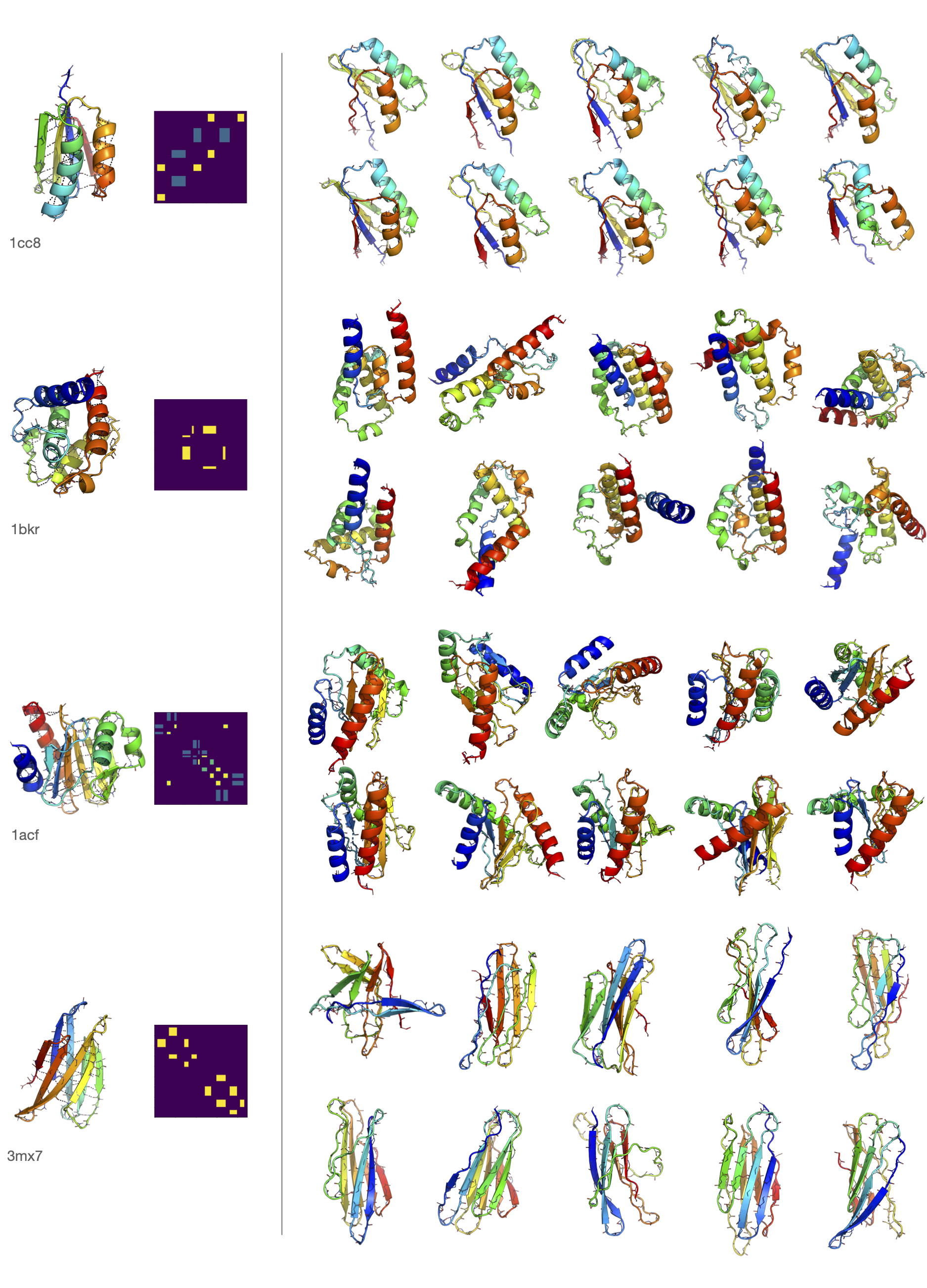}
  \caption{Random generated backbone samples. }
\label{fig:random_backbones}
\end{figure}

\begin{figure}[ht]
  \centering
    \includegraphics[width=0.95\textwidth]{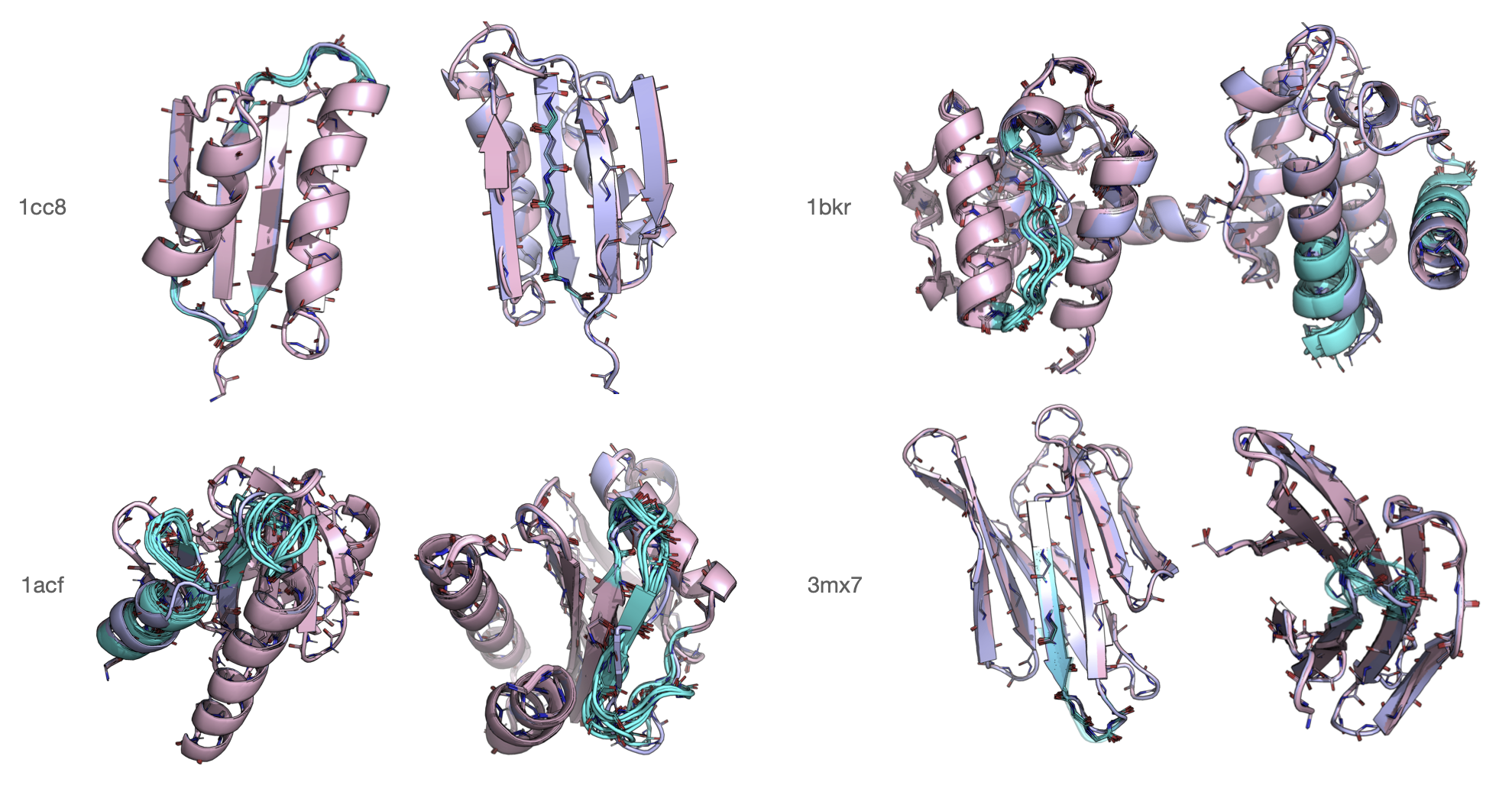}
  \caption{Structure modification: Random generated inpainting samples.}
\label{fig:random_inpaint}
\end{figure}

\begin{figure}[ht]
  \centering
    \includegraphics[width=0.95\textwidth]{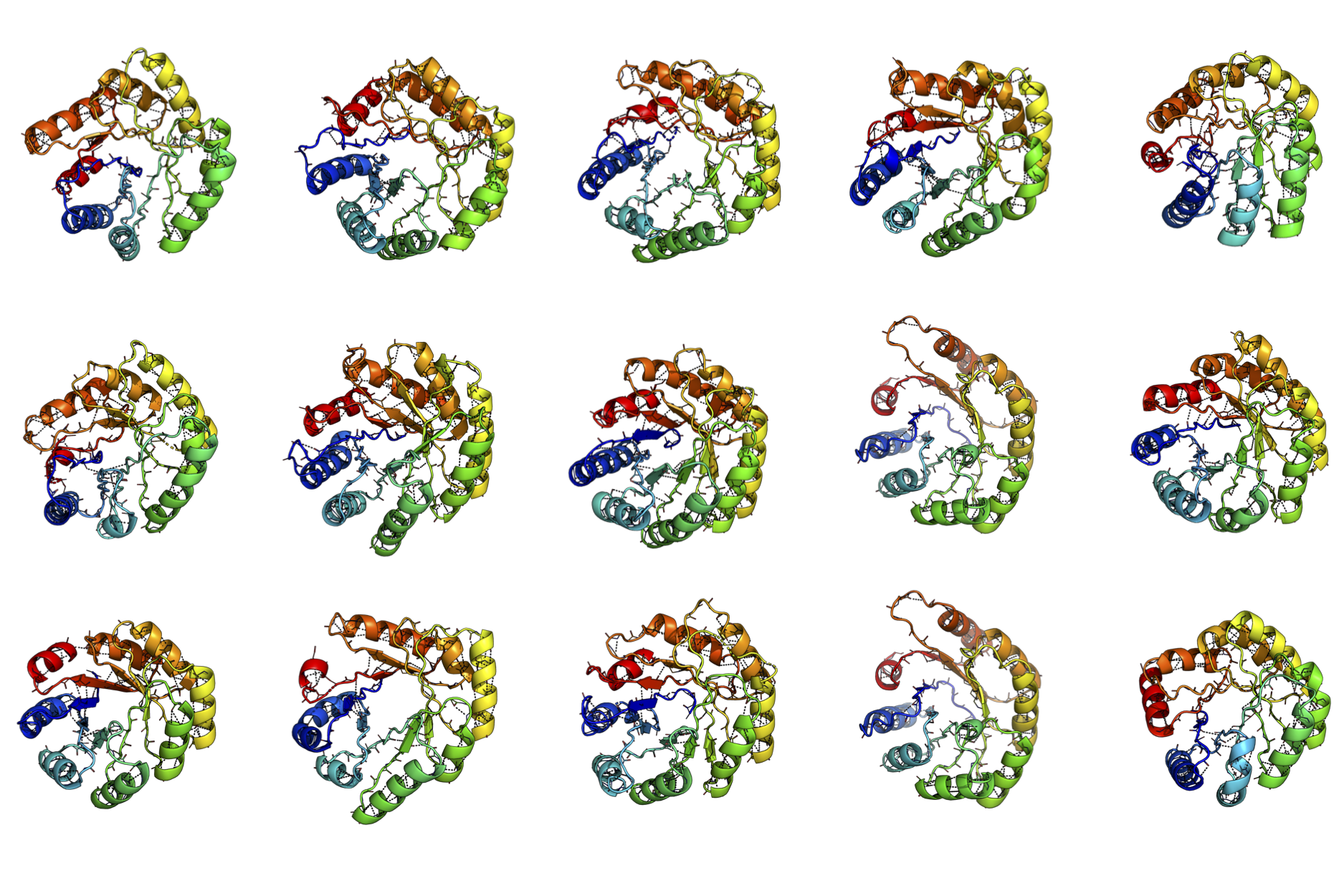}
  \caption{Random generated TIM-barrels with varied underlying four-fold symmetric topologies. Lengths of secondary structure elements in each quarter are sampled randomly. }
\label{fig:random_TIMs}
\end{figure}

\begin{figure}[ht]
  \centering
    \includegraphics[width=0.95\textwidth]{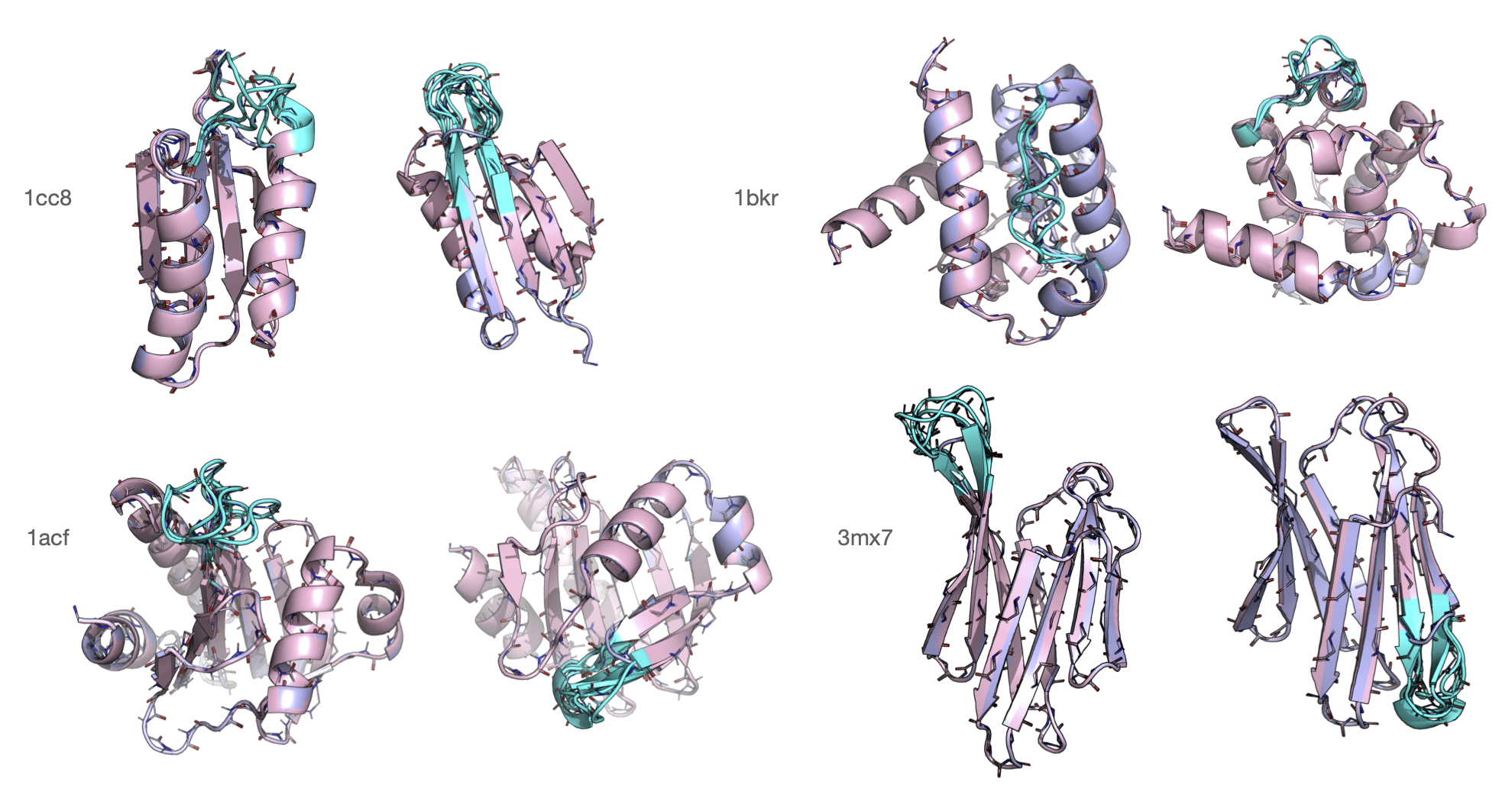}
  \caption{Structure modification: Sampling loops of varying lengths.}
\label{fig:modify_loop_len}
\end{figure}

\end{document}